\documentstyle[psfig]{elsart}
\begin{document}

\begin{frontmatter}

\title{Correlation functions and emission time sequence \\
of light charged particles from projectile-like fragment source \\
in $E/A$ = 44 and 77 MeV $^{40}$Ar + $^{27}$Al collisions
}

\author{
R.~Ghetti$^{a,1}$, and J.~Helgesson$^{b}$
}
\address{$^a$Department of Physics, Lund University, Box 118, SE-221 00 Lund, Sweden\\
$^b$School of Technology and Society, Malm\"o University, SE-205 06 Malm\"o, Sweden
}
\author{
G.~Lanzan\`o$^{c}$, 
E.~De Filippo$^{c}$,
M.~Geraci$^{c}$,
S.~Aiello$^{c}$,
}
\author{
S.~Cavallaro$^{d}$,
A.~Pagano$^{c}$,
and G.~Politi$^{c}$ 
}
\address{
$^c$Dipartimento di Fisica dell'Universit\`a di Catania, 
Istituto Nazionale di Fisica Nucleare, Sezione di Catania, 
Via S.\ Sofia 64, I-95123 Catania, Italy \\
$^d$Laboratori Nazionali del Sud, Catania, 
and Dipartimento di Fisica dell'Universit\`a di Catania, 
Via S.\ Sofia 44, I-95123 Catania, Italy 
}
\author{
J.~L.~Charvet$^{e}$,
R.~Dayras$^{e}$,
E.~Pollacco$^{e}$, and 
C.~Volant$^{e}$
}
\address{
 $^e$DAPNIA/SPhN, CEA/Saclay, F-91191 Gif-sur-Yvette CEDEX, France
}
\author{
C.~Beck$^{f}$,
D.~Mahboub$^{f,+}$,
and R.~Nouicer$^{f,++}$
}
\address{
$^f$Institut de Recherches Subatomiques, UMR7500, CNRS-IN2P3 and 
Universit{\'e} Louis Pasteur, B.P.\ 28, F-67037 Strasbourg CEDEX 2, France 
}

\begin{abstract}
Two-particle correlation functions, 
involving protons, deuterons, tritons, and $\alpha$-particles, 
have been measured at very forward angles 
(0.7$^{\rm o} \le \theta_{lab} \le 7^{\rm o}$),  
in order to study projectile-like fragment (PLF) emission in 
$E/A$ = 44 and 77 MeV $^{40}$Ar + $^{27}$Al collisions. 
Peaks, originating from resonance decays, are larger at 
$E/A$ = 44 than at 77 MeV. 
This reflects the larger relative importance of independently 
emitted light particles, as compared to two-particle decay 
from unstable fragments, at the higher beam energy. 
The time sequence of the light charged particles, emitted 
from the PLF, has been deduced from particle-velocity-gated 
correlation functions (discarding the contribution from 
resonance decays). 
$\alpha$-particles are found to have an average emission time 
shorter than protons but longer than tritons and deuterons.
\end{abstract}

\date{\today}

\end{frontmatter}

+) Present address: University of Surrey, Guildford, Surrey GU2 7XH, UK.\\
++) Present address: Brookhaven Nat.\ Lab., Upton, New York, 11973-5000, USA.\\

PACS number(s): 25.70.Pq, 25.70.Mn \\


%Preprint number: nucl-ex/XXXX \\ 

{\noindent \small \rm $^1$Corresponding author. 
Department of Physics, University of Lund. \\ 
Box 118, S-22100 Lund, Sweden. Tel. +46-(0)46-2227647. \\ 
E-mail address: roberta.ghetti@nuclear.lu.se} (R.\ Ghetti).

\maketitle

%%%%%%%%%%%%%%%%%%%%%%%%%%%%%%%%%%%%%%%%%%%%%%%%%%%%%%%%%%%%%%%%%%%%%%%%
\section{Introduction}
\label{sec:intro}

Extensive studies have demonstrated that semiperipheral heavy-ion 
collisions at intermediate energies proceed through a dissipative 
binary reaction mechanism, characterized by early dynamical emission 
from an intermediate velocity source, followed by statistical 
evaporation from an excited projectile-like fragment and from 
an excited target-like fragment 
\cite{Schr92,Boug95,Laro95,Rive96,Leco96,Skul96,Dorv99}.
In particular, this is true for the reverse kinematics reaction 
Ar + Al, investigated by several authors 
\cite{Dayras86,Dayras89,Hagel89,Pete90,Pete95,Eudes,Angelique,PRL-03,Volo04,Lanza98,Lanza01}. 
In this paper, we aim to study light charged particle (LCP) 
emission from the PLF, in $E/A$ = 44 and 77 MeV $^{40}$Ar + $^{27}$Al collisions. 
To this end, we investigate experimental two-particle correlation functions 
of protons, deuterons, tritons and $\alpha$-particles. 

From previous investigations, as well as from kinematical reasons, 
it is well known that semiperipheral collisions are dominated by PLF 
emission at very forward angles (see, e.g., the study of 
LCP emission from $E/A$ = 60 MeV  $^{40}$Ar + $^{27}$Al collisions 
presented in Ref.\ \cite{Lanza98}).
In the present study, the experimental conditions of low detection 
thresholds, large dynamic range, fine granularity, and small angular 
separation between the centers of adjacent detectors, allows us to 
perform the correlation function analysis in a very forward 
angular range, 0.7$^{\rm o} \le \theta_{lab} \le 7^{\rm o}$.

Information about the emission times of the different particle types 
can be deduced from a model-independent analysis of the respective 
measured correlation functions. 
In particular, we aim to understand the order of emission 
of the LCPs from the PLF, by means of particle-velocity-gated 
correlation functions of non-identical particles \cite{Lednicky,PRL-01,Gourio}. 
This adds an important and novel piece of information 
to the picture of the reaction mechanism emerging from 
previous studies \cite{Lanza98,Lanza01}.

Furthermore, we investigate the relative importance of independent 
(two-) LCP emission from the PLF, as compared to two-particle decay 
from unstable fragments emitted in the reaction. This is 
accomplished by comparing the strength of the corresponding 
resonance peaks in the correlation function at the two beam energies. 

The paper is organized as follows. 
After a brief description of the experimental setup 
and event selection criteria (Sec.\ \ref{sec:exp}), 
we present the results from the data analysis of the 
$E/A$ = 44 and 77 MeV $^{40}$Ar + $^{27}$Al collisions. 
These include LCPs kinetic energy spectra 
(Sec.\ \ref{sec:energy}), correlation functions of identical 
and non-identical particle pairs (Sec.\ \ref{sec:cf}), 
and particle emission time sequence extracted from the 
velocity-gated two-particle correlation functions 
(Sec.\ \ref{sec:lednicky}). 
Finally, a summary and conclusions are given in 
Sec.\ \ref{sec:summary}.

%%%%%%%%%%%%%%%%%%%%%%%%%%%%%%%%%%%%%%%%%%%%%%%%%%%%%%%%%%%%%%%%%%%%%%%
\section {Experimental details}
\label{sec:exp}

The data analyzed in this paper were taken at GANIL 
in two separate experiments with different beam energies.
$E/A$ = 44 and 77 MeV $^{40}$Ar pulsed beams impinged 
on 200 $\mu$g/cm$^2$ thick $^{27}$Al targets. 
The reaction products were detected by the ARGOS \cite{Lanza01} 
multidetector system, an array of 112 hexagonal BaF$_2$ crystals 
(surface area 25 cm$^2$) modified into phoswich by 
means of fast plastic scintillator sheets (700 and 1900 $\mu$m thick 
for the two beam energies, respectively).

ARGOS was placed inside the Nautilus vacuum chamber, with the following 
geometry (which is illustrated in Fig.\ 1 of Ref.\ \cite{Lanza01}). 
A forward wall of 60 elements was 
placed between 0.7$^{\rm o}$ and 7$^{\rm o}$ in a honeycomb shape at a 
distance of 233 cm from the target (solid angle 0.03 sr).
The angular separation between the centers of two adjacent detectors 
was $\approx$ 1.5$^{\rm o}$. 
A backward wall of 18 phoswich elements 
was placed between  160$^{\rm o}$ and 175$^{\rm o}$, 
at a distance of 50 cm from the target. 
Finally, a battery of 30 phoswich elements was placed in plane, 
at angles between 10$^{\rm o}$ and 150$^{\rm o}$, 
at a distance from the target variable from 50 to 200 cm. 
Further details on the experimental setup can be found in 
Refs.\ \cite{Lanza01,LanzaNIM,Mario}.

Identification of the reaction products was achieved via 
shape discrimination of the photomultiplier signals and 
Time-of-Flight technique \cite{Lanza01,LanzaNIM}. 
LCPs ($Z$ = 1 and 2) were isotopically separated, while 
fragments with $Z \ge$ 3 were identified in charge. 
The particle velocity was directly measured by means of 
Time-of-Flight technique. The experimental thresholds were 
set to 2 cm/ns for $E/A$ = 44 MeV, and to 3 cm/ns 
for $E/A$ = 77 MeV. 
The BaF$_2$ crystals, of variable thickness up to 10 cm, 
stopped protons of energy up to 200 MeV. 

Two slightly different trigger conditions were implemented 
in the two experiments. In the 44 MeV experiment, 
events were recorded each time the in-plane detectors or the 
backward wall was fired, a minimum total multiplicity of 
2 being registered. 
In the 77 MeV experiment, instead, a coincidence between any 
two detectors was sufficient to start the data acquisition. 

In this analysis, the event selection aims to choose semiperipheral 
collisions. This is achieved by requiring that the pairs of LCPs 
($p$, $d$, $t$, and $\alpha$) detected in the forward wall, 
must be in coincidence with one PLF 
(any fragment with atomic charge 3~$\le Z_{PLF} \le$~18 
and velocity within 70$\%$ of the beam velocity)
also detected in the forward wall\footnote{For the $E/A$ = 44 MeV 
experiment, an additional particle detected in the plane or in the 
backward detectors is required, since this condition was implemented 
in the hardware trigger of that experiment.}.
The analysis of particle velocity spectra and of invariant 
cross-sections presented in Refs.\ \cite{Lanza01,Mario}, 
indicates that this type of events is associated 
with semiperipheral binary collisions, where two sources, with 
velocities close to the initial velocity of the projectile and 
of the target nuclei, are formed. In addition, the mid-rapidity 
region is an abundant source of particles \cite{Lanza01}. 

%%%%%%%%%%%%%%%%%%%%%%%%%%%%%%%%%%%%%%%%%%%%%%%%%%%%%%%%%%%%%%%%%%%%%%%
\section {Kinetic energy distributions of light charged particles}
\label{sec:energy}

\begin{figure}
\centerline{\psfig{file=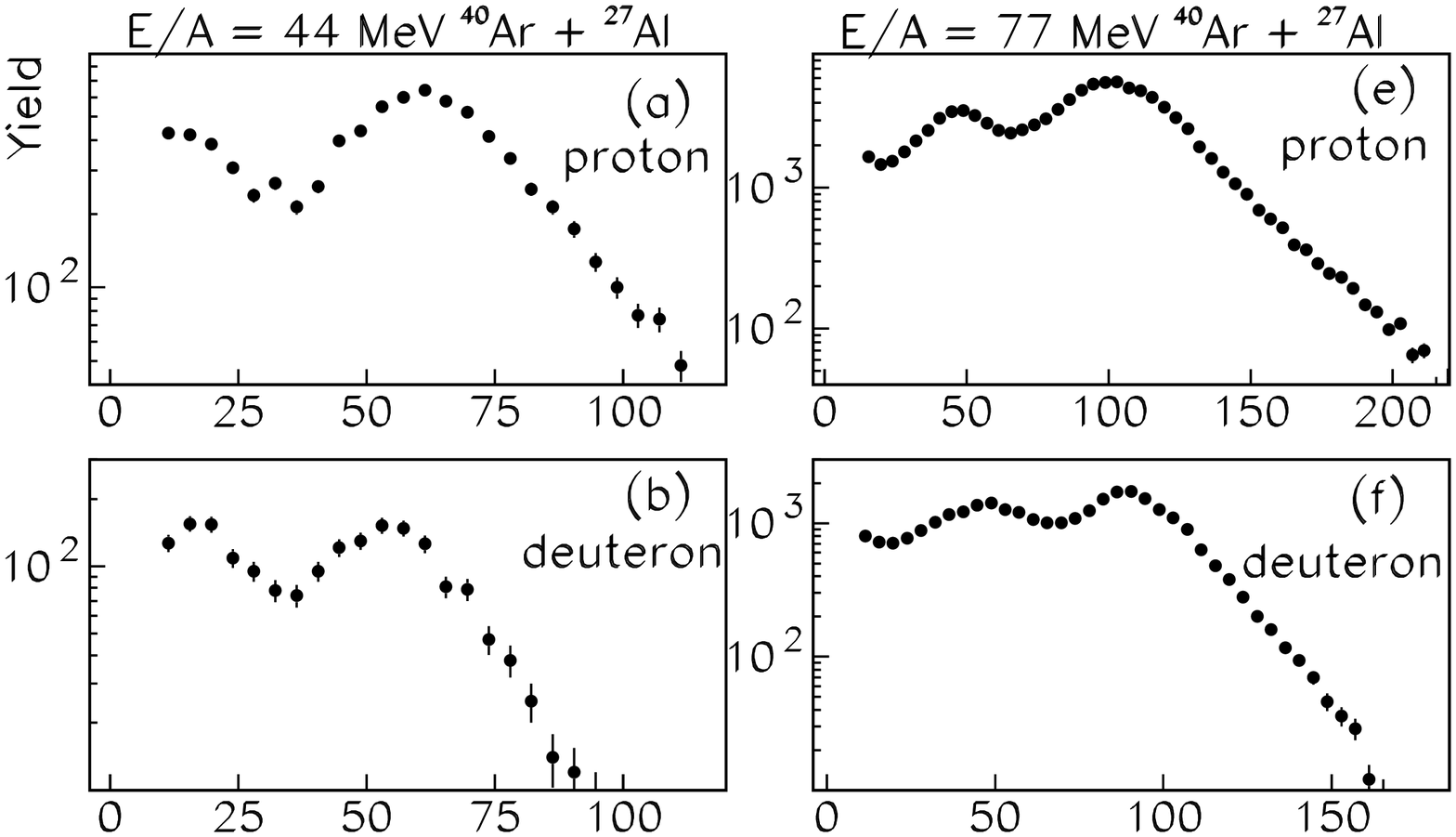,height=9cm,angle=0}}
\centerline{\psfig{file=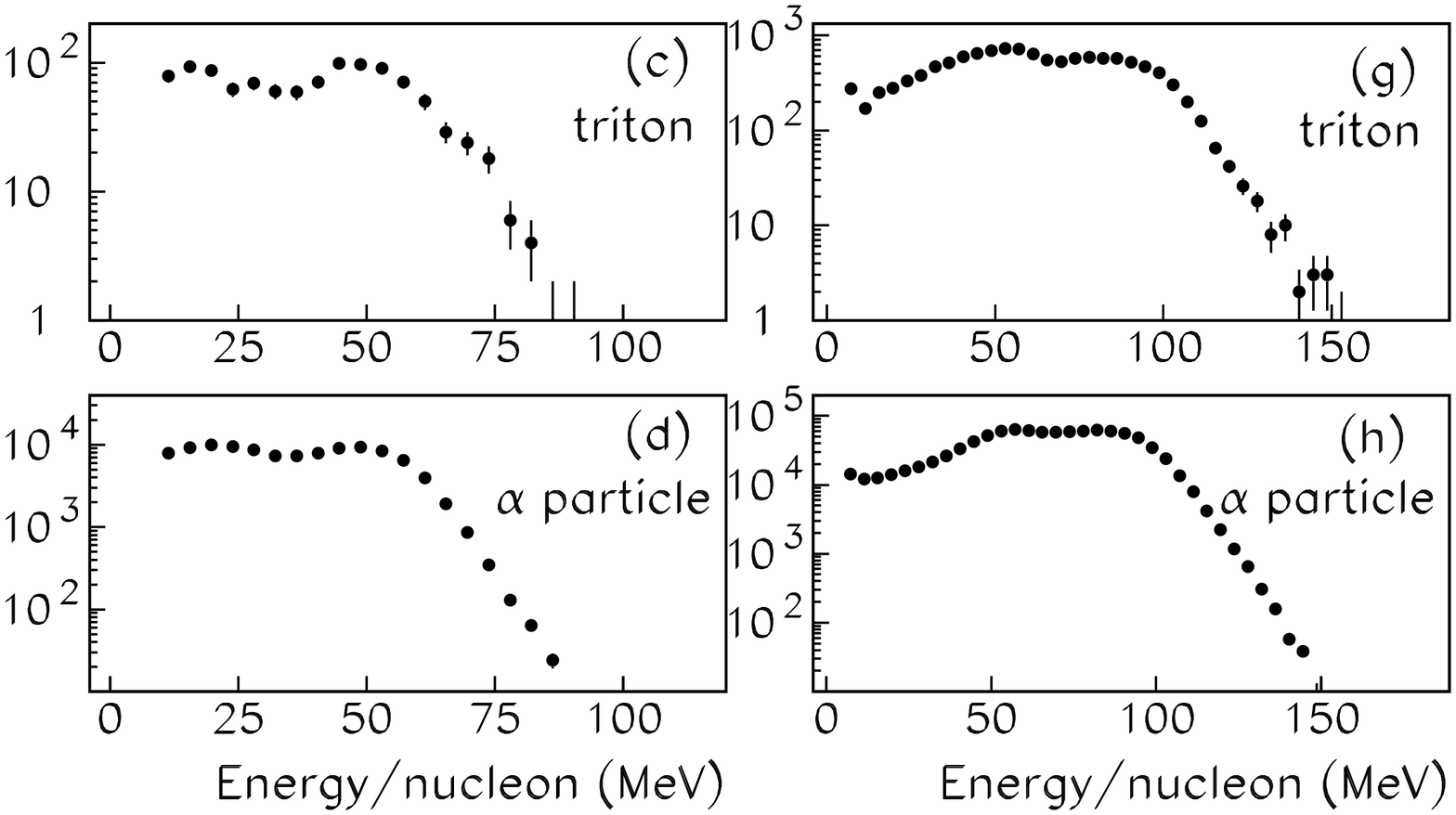,height=9cm,angle=0}}
\caption{
From $E/A$ = 44 MeV (left column) and 77 MeV (right column) 
$^{40}$Ar + $^{27}$Al collisions, 
distributions of the laboratory kinetic energy/nucleon of the 
LCPs detected at 0.7$^{\rm o} \le \theta_{lab} \le 7^{\rm o}$. 
At a given bombarding energy, all the spectra have the same 
(arbitrary) normalization.
}
\label{spectra}
\end{figure}

The laboratory kinetic energy distributions of protons, deuterons, tritons, 
and $\alpha$-particles, 
measured in the $E/A$ = 44 and 77 MeV $^{40}$Ar + $^{27}$Al reactions, 
are shown in Fig.\ \ref{spectra}, integrated in the 
angular range 0.7$^{\rm o} \le \theta_{lab} \le 7^{\rm o}$. 
The detection energy thresholds correspond to $\approx$ 2 and 5 MeV/nucleon 
for $E/A$ = 44 and 77 MeV experiment, respectively. 

One can notice several qualitative features: 

1) The spectra are dominated by a dumb-bell structure, 
displaying a minimum slightly below the beam energy. 
This behavior is typical of those particles (from the 
isotropically emitting PLF) that have a velocity 
component either parallel to PLF movement 
(higher energy peak) or antiparallel (lower energy peak) \cite{Lanza01}. 
Because of the Coulomb push from the PLF, these particles 
will have an energy either higher or lower than the beam energy 
\cite{Brief04}.

2) In addition to PLF emission, the spectra also contain contributions 
from an intermediate velocity source and from a target-like 
source \cite{Lanza01}. 

3) The  dumb-bell structure in the energy spectra is more pronounced 
for $p$ and $d$ and it is slightly suppressed for $t$ and $\alpha$. 
This may be expected, since the Coulomb push per particle 
experienced by an $\alpha$-particle is less than half that 
experienced by a proton 
(taking into account the mass and charge of the $\alpha$-particle, 
and its larger average emission distance due to its larger size). 
In addition, the kinematical focusing at forward angles increases 
with the mass of the emitted particles, and it is strongest for 
$\alpha$-particles \cite{Lanza98}. Another effect for the smearing  
the dumb-bell structure may come from recoil effects. Finally, 
a reduced dumb-bell shape may be caused by an enhanced contribution 
of non-PLF decay mechanisms (such as intermediate velocity source 
emission and secondary emission from the decay of highly excited 
primary fragments) for $t$ and $\alpha$ emission.
Indeed, the average relative intensity of intermediate velocity 
source emission is larger for $\alpha$-particles than for protons, 
as demonstrated in Ref.\ \cite{Lanza01} for the $E/A$ = 44 MeV 
$^{40}$Ar + $^{27}$Al reaction. 

4) The dumb-bell shape associated with PLF decay is slightly 
more pronounced at $E/A$ = 44 MeV than at 77 MeV. 
This is partly explained by the fact that the 
Coulomb push becomes relatively less important 
at 77 MeV, as compared to the boost from the source 
system to the laboratory system. In addition, 
it may be partly explained by an 
enhanced contribution of non-PLF decay mechanisms at the 
higher energy (intermediate velocity source emission and 
secondary decay) \cite{Lanza01,Brief04}.

5) The yield of $\alpha$-particles is rather large 
in this forward angular range.
Speculations on the origin of these copious 
$\alpha$-particles have been discussed in  Refs.\ \cite{Lanza98,Lanza01}. 
It could be due to the production in the reaction of excited 
light ions, and/or to the $\alpha$ structure of the two interacting 
nuclei \cite{Hodgson}. 

%%%%%%%%%%%%%%%%%%%%%%%%%%%%%%%%%%%%%%%%%%%%%%%%%%%%%%%%%%%%%%%%%%%%%%%%%%%
\section{Two-particle correlation functions}
\label{sec:cf}

The experimental correlation function 
is constructed by dividing the coincidence yield, $N_c$, by the yield 
of uncorrelated events, $N_{nc}$, 
\begin{equation}
C(\vec{q},\vec{P_{tot}}) = k \frac {N_c(\vec{q},\vec{P_{tot}}) }
{ N_{nc}(\vec{q},\vec{P_{tot}})}.
\end{equation}
$\vec{q} = \mu (\vec{p}_{1} / m_1 - \vec{p}_{2} / m_2)$ 
is the relative momentum, $\mu$ is the reduced mass, 
and $\vec{P_{tot}} = \vec{p_1} + \vec{p_2}$ 
is the total momentum of the particle pair.
In the following, an implicit integration over the six variables 
$\vec{q}$ and $\vec{P_{tot}}$ is performed. 
The correlation function is normalized to unity at large values of $q$, 
where no correlations are expected.
The background yield, $N_{nc}$, can be constructed either from the product 
of the singles distributions  \cite{Singles} 
or with the so-called ``event-mixing'' 
technique \cite{Kopylov}, combining particles from different events. 
We have verified that, for the present data set, 
the two techniques yield the same results, within the 
experimental uncertainty. Thus, we have chosen to utilize the 
singles technique.

In the present analysis, only particles with energy/nucleon 
larger than a value slightly below the beam energy have been 
utilized to construct $C(q)$. This is because 
combining particles from both peaks of the dumb-bell shaped energy 
distributions (Fig.\ \ref{spectra}), 
introduces correlations other than two-body final state interactions
and/or quantum symmetrization effects. These unwanted correlations 
most likely originate from a mixing of sources with different 
velocities. For more details, see Ref.\ \cite{Brief04}. 
By applying this energy cut, the contribution from the PLF 
source relative to other sources is enhanced, while the 
statistics and the accessible range of relative momenta 
are reduced. Nevertheless, thanks to the small angular separation 
between the centers of two adjacent detectors ($\approx$ 1.5$^{\rm o}$), 
the region of small relative momenta (where interesting signatures of 
final state interactions and antisymmetrization effects appear) 
is still populated.

Before we proceed to analyze the experimental results, we would 
like to point out that, for those correlation functions  
characterized by final state interaction leading to resonances, 
the resonance peaks may a priori have two different origins 
\cite{Pochodzalla,Deyoung}:

i) From processes where an unstable fragment formed in the reaction 
decays into the two measured particles.
If the secondary decay of a specific fragment 
produces two particles that are measured in coincidence 
(like in e.g.\ $^8$Be $\rightarrow \alpha + \alpha$ 
or $^6$Li $\rightarrow$ $\alpha$ + $d$),
the decay will give rise to a resonance peak in 
the corresponding correlation function.

ii) From interactions between independently emitted particles. 
De-excitation of the PLF and secondary decay of excited fragments 
may produce particles are measured in coincidence.
The relative distance for most such pairs 
will be large (in both space and time) and they 
will interact weakly. This means that they will give 
rise to correlation effects mainly at small relative 
momenta \cite{Verde}. 

Figure \ref{npa} presents the correlation functions 
of particle pairs measured in the 
$E/A$ = 44 and 77 MeV $^{40}$Ar + $^{27}$Al experiments. 
The events are selected with the conditions described in Sec.\ \ref{sec:exp}.
The low energy thresholds are set to 35 MeV/nucleon for the $E/A$ = 44 MeV data, 
and 65 MeV/nucleon for the $E/A$ = 77 MeV data, 
for the reasons discussed above and in Ref.\ \cite{Brief04}. 

\subsection{The $pp$ correlation function}
The $pp$ correlation function 
(normalized in the relative momentum region $q$ = 80--100 MeV/c) 
is shown in Figs.\ \ref{npa}(a,f). 
This result was already presented in Ref.\ \cite{Brief04}. 
It exhibits a shallow maximum at $q \approx$ 20 MeV/c, 
caused by the attractive $s$-wave interaction 
between the two protons.
The maximum at 20 MeV/c may be seen as the 
resonance peak from the decay of the particle unstable $^2$He. 
The minimum at $q \rightarrow$ 0 MeV/c is due to the interplay 
between quantum antisymmetrization effects and final state 
Coulomb repulsion between the two protons \cite{Koonin}.
The very weak peak at 20 MeV/c indicates that proton emission occurs 
on a long time scale \cite{Koonin}.

\subsection{The $p\alpha$ correlation function}
The $p\alpha$ correlation function (normalized 
in the relative momentum region $q$ = 100--150 MeV/c) 
is shown in Figs.\ \ref{npa}(b,g). 
The broad peak at $q \approx$ 54 MeV/c is due 
to the unbound ground state of $^5$Li 
(1.69 MeV above the $p$ + $\alpha$ threshold, 
$\Gamma_{cm}$ = 1.23 MeV) \cite{Tilley1}. 
Most probably, the bump contains additional contributions 
from a background of $p$ and $\alpha$-particles emitted 
independently from source de-excitation 
(the large yields of $p$ and $\alpha$ \cite{Lanza98,Lanza01} 
enhance the probability that two such particles are emitted 
close enough in time that they will experience strong 
final state interaction). 

The very similar behavior of the two $pp$ and $p\alpha$ correlation 
functions at $E/A$ = 44 and 77 MeV, indicates that the characteristics 
of PLF proton and $\alpha$-particle emission are rather independent 
of the beam energy. 
As demonstrated by the statistical calculations of Ref.\ \cite{Hudan}, 
different emitted particles from the PLF are sensitive to different 
portions of the PLF de-excitation cascade. 
Because of the Coulomb barrier and of the binding energy relative to 
the available excitation of the PLF, the emission time distribution 
of protons and $\alpha$-particles is expected to be rather flat. 
In the calculations of \cite{Hudan}, the emission time distribution 
exhibited by $\alpha$-particles is found to decrease by only a factor 
of 3 over the time interval of 150 fm/c, while other particles (e.g.\ 
tritons) exhibit steeper distributions, decreasing more than one order 
of magnitude in the same time interval. 
Thus, protons and $\alpha$-particles are expected to sample the end of PLF 
de-excitation cascade, and this effect is expected to be approximately the 
same at 44 and at 77 A MeV. In addition, there might be a large background 
of independently emitted protons and $\alpha$-particles from secondary 
decay of excited fragments, also probing the long time scale.

\subsection{The $d\alpha$ correlation function}
\begin{figure}
\centerline{\psfig{file=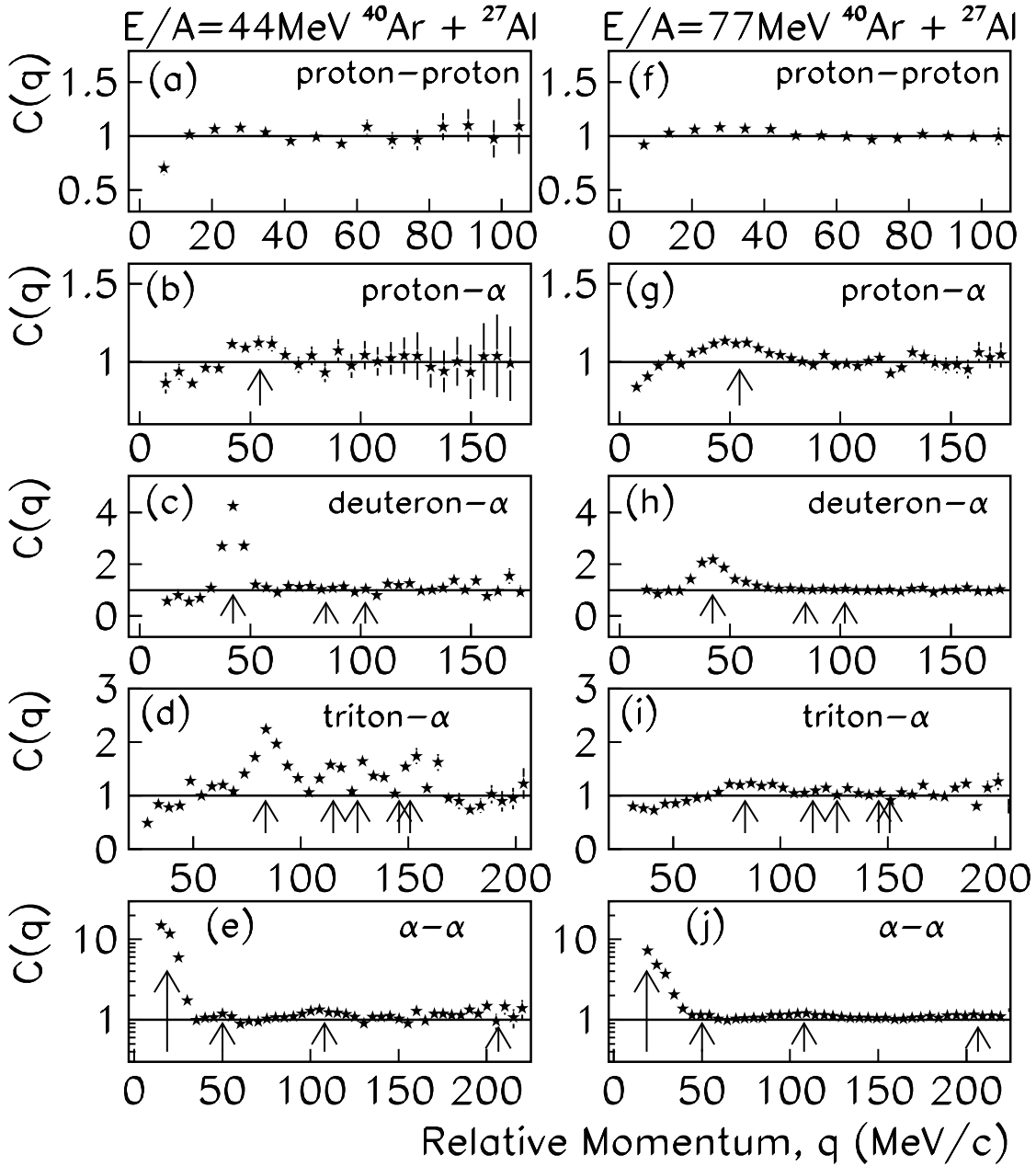,height=18.cm,angle=0}}
\caption{
From $E/A$ = 44 (left column) and 77 (right column) MeV 
$^{40}$Ar + $^{27}$Al collisions,
$pp$ [panels (a,f)], 
$p\alpha$ [panels (b,g)], 
$d\alpha$ [panels (c,h)], 
$t\alpha$ [panels (d,i)], 
and $\alpha\alpha$ [panels (e,j)]
correlation functions. The position of 
expected resonance peaks is indicated 
by the arrows.
}
\label{npa}
\end{figure}

The $d\alpha$ correlation function (normalized in the relative 
momentum region $q$ = 110--150 MeV/c) is shown in 
Figs.\ \ref{npa}(c,h). 
It is governed by the strong resonance at 
$q \approx$ 42.2 MeV/c, due to the 2.186 MeV 
($\Gamma_{cm}$ =0.024 MeV) excited state of $^6$Li \cite{Tilley1}.
This resonance peak reaches the value 
$C(q \approx 42) \approx$ 4.2 at $E/A$ = 44 MeV, 
as compared to $C(q \approx 42) \approx$ 2.2  
at $E/A$ = 77 MeV. 
The resonances expected at $q \approx$ 84 MeV/c and 102 MeV/c 
(from the overlap of the $^6$Li excited states 
at 4.31 MeV, $\Gamma_{cm}$ = 1.30 MeV, and 5.65 MeV, 
$\Gamma_{cm}$ = 1.50 MeV) \cite{Tilley1} are hardly seen in the data. 

\subsection{The $t\alpha$ correlation function}
The $t\alpha$ correlation function (normalized in the relative 
momentum region $q$ = 180--200 MeV/c) is shown in 
Figs.\ \ref{npa}(d,i). Because of experimental difficulties 
in the identification of low energy tritons and $\alpha$-particles, 
the $t\alpha$ correlation function is presented in the region  
$q >$ 30 MeV/c [the x-axis scale is zero-suppressed in 
Figs.\ \ref{npa}(d,i)]. The $t\alpha$ correlation function is 
characterized by a maximum at $q \approx$ 83.6 MeV/c, 
due to the 4.652 MeV ($\Gamma_{cm}$ =0.069 MeV) 
excited state of $^7$Li \cite{Tilley1}.
This resonance peak reaches the value 
$C(q \approx 83) \approx$ 2.1 at $E/A$ = 44 MeV, 
as compared to $C(q \approx 83) \approx$ 1.2  
at $E/A$ = 77 MeV. 
Other excited states of $^7$Li that decay into 
$t\alpha$ are at 
6.60 MeV ($\Gamma_{cm}$ =0.918 MeV), 
7.45 MeV ($\Gamma_{cm}$ =0.080 MeV), 
9.09 MeV ($\Gamma_{cm}$ =2.752 MeV), and 
9.57 MeV ($\Gamma_{cm}$ =0.437 MeV) \cite{Tilley1}. 
The corresponding resonances, expected at 
$q$ = 115.1, 126.3, 145.7, and 150.8 MeV/c,
are visible in the $E/A$ = 44 MeV data. 

\subsection{The $\alpha\alpha$ correlation function}
The $\alpha\alpha$  correlation function, 
(normalized in the relative 
momentum region $q$ = 140--180 MeV/c)
is shown in Figs.\ \ref{npa}(e,j), in semi-logarithmic scale. 
This result was already presented in Ref.\ \cite{Brief04}. 
This correlation function is dominated by the decay of 
the particle unstable ground state of 
$^8$Be ($\Gamma_{cm}$ = 5.57 eV) at $q$ = 18.6 MeV/c \cite{Tilley2}.
This resonance peak reaches the value 
$C(q \approx 18) \approx$ 13.5 at $E/A$ = 44 MeV, 
and $C(q \approx 18) \approx$ 8.5 at $E/A$ = 77 MeV.
The decay of the 2.43 MeV excited state of $^9$Be ($\Gamma_{cm}$ = 0.78 keV) 
into the undetected neutron and the unstable $^8$Be ground-state, 
as well as the decay of the 3.03 MeV ($\Gamma_{cm}$ = 1.51 MeV) 
and 11.35 MeV ($\Gamma_{cm}$ = 3.5 MeV) excited states of $^8$Be \cite{Tilley2}, 
lead to resonance peaks at $q \approx$ 50, 108, and 206 MeV/c, more clearly 
visible in the $E/A$ = 44 MeV data. 

In summary, comparing the results from the two beam energies, 
we have seen that the resonance peaks are more pronounced at 
$E/A$ = 44 MeV than at 77 MeV. 
The effect is largest for the $d\alpha$ correlation function 
(ratio of main peaks is 1.90), 
followed by the $t\alpha$ (ratio of main peaks is 1.75) and 
the $\alpha\alpha$ (ratio of main peaks is 1.59). 
This indicates that at $E/A$ = 77 MeV there is a larger yield 
of uncorrelated light particles (in particular $d$ and $t$) 
as compared to heavy fragments 
(in particular $^8$Be, $^7$Li, and $^6$Li). 
This finding is in agreement with the results from the 
analysis of the particle-velocity-gated correlation functions, 
presented in the next section.

%%%%%%%%%%%%%%%%%%%%%%%%%%%%%%%%%%%%%%%%%%%%%%%%%%%%%%%%%%%%%%%%%%%%%%%%%%%
\section{Particle emission time sequence}
\label{sec:lednicky}

Correlation functions of non-identical particles can be utilized to 
provide information about the order of emission of the 
different particles \cite{Kotte}. To this end, particle-velocity-gated 
correlation functions are constructed \cite{Lednicky,PRL-01}.

For non-identical particles, say $a$ and $b$, we construct 
the correlation functions $C_a(q)$, gated on pairs 
with $v_a > v_b$, and $C_b(q)$, gated on pairs 
with $v_b > v_a$. If particle $a$ is emitted later (earlier) 
than particle $b$, the ratio $C_a/C_b$ will show a peak 
(dip) in the region of $q$ where there is a correlation, 
and a dip (peak) in the region of $q$ where there is an 
anticorrelation. The particle velocities are calculated 
in the frame of the PLF. A single normalization constant, 
calculated from the ungated correlation function, is utilized 
for both $C_a$ and $C_b$ \cite{PRL-01}.

The $p\alpha$, $d\alpha$, and $t\alpha$ correlation functions 
are characterized by final state interaction leading to resonances. 
If the two particles are emitted independently, the velocity-gated 
correlation functions contain information on the time sequence, 
but this is not the case if the two particles originate from the 
decay of an unstable fragment emitted in the reaction. 
In the latter case, the two 
particle velocities are determined by momentum conservation, 
and, in the rest system of the decaying fragment, the lightest 
particle will always get the highest velocity. 
Therefore, the expected behavior for particle pairs coming 
from a two-body decay of a decaying fragment is the following: 
the velocity-gated correlation function, obtained with 
the condition that the lightest particle has the 
largest velocity, should exhibit the strongest correlation 
or anticorrelation.
Therefore, when we, in some cases, observe the opposite behavior 
(namely that the gate where {\it the heaviest} particle has the largest 
velocity leads to a stronger correlation or anticorrelation) 
we can reliably conclude that this behavior is dominated by 
a mechanism different than two-body decay. We attribute this 
effect to the interaction of independently emitted particles, 
and, in this case, we use the velocity-gated correlation 
function to obtain information on the time sequence of the 
independently emitted particles \cite{Helgesson}.

\subsection{The particle-velocity-gated $p\alpha$ correlation function}
\begin{figure}
\vspace{0.4cm}
\centerline{\psfig{file=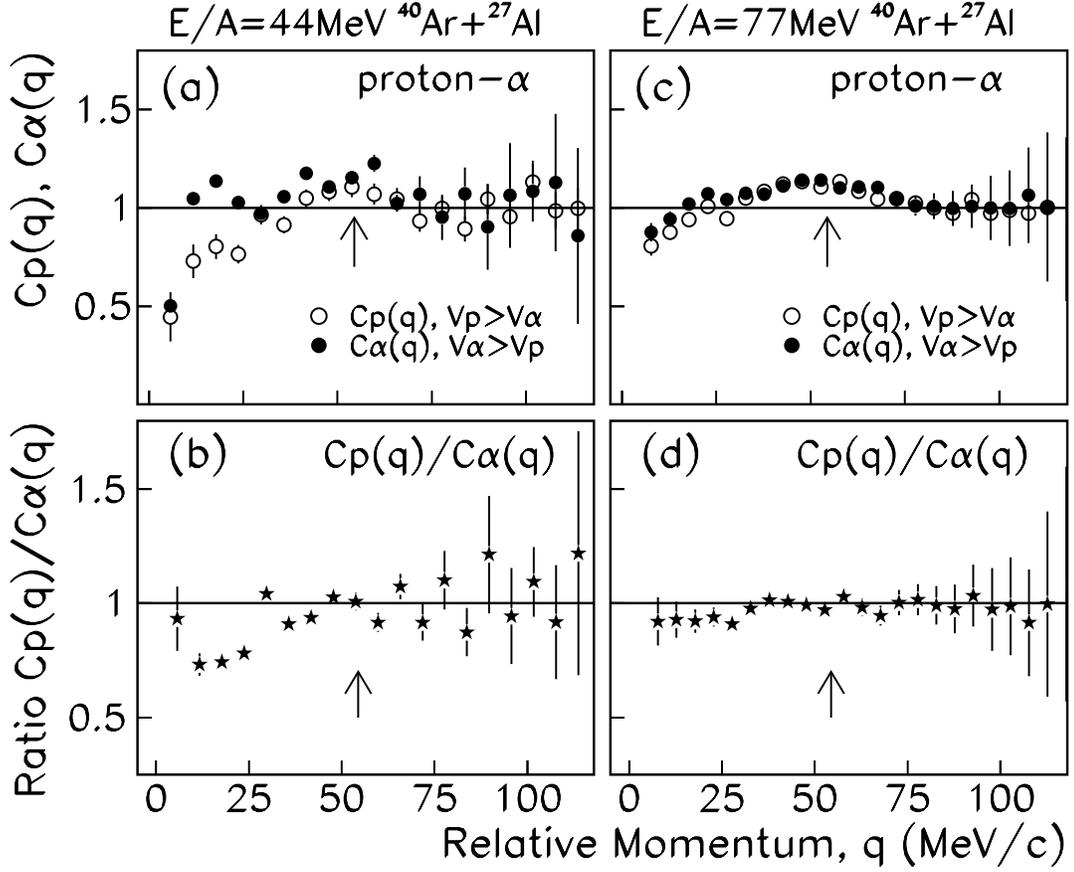,height=12.5cm,angle=0}}
\caption{
From $E/A$ = 44 MeV (left column) and 77 MeV (right column) 
$^{40}$Ar + $^{27}$Al collisions, 
upper panels: particle-velocity-gated 
(filled and open circles) $p\alpha$ correlation functions; 
lower panels: the ratio of the particle-velocity-gated 
correlation functions. 
}
\label{ledpa}
\end{figure}

The particle-velocity-gated $p\alpha$ correlation function 
is shown in Fig.\ \ref{ledpa}. 
In the relative momentum region dominated by the two-body 
decay of $^5$Li ($q \approx$ 54 MeV/c) the enhancement of 
$C_p$ ($v_p > v_\alpha$) over $C_\alpha$ ($v_\alpha > v_p$), 
expected from momentum conservation, is not observed. 
This is a first indication that the behavior of the $p\alpha$ 
correlation function is dominated by a background of independently 
emitted pairs.

The emission chronology can be deduced from the behavior 
at $q <$ 30 MeV/c, where there are no resonant 
states that directly lead to the formation of $p\alpha$ pairs, and 
where the $p\alpha$ correlation function displays an anticorrelation 
(\cite{Boal2} and Fig.\ 2). 
The fact that pairs with $v_p > v_\alpha$ (contributing to $C_p$) 
interact more strongly (stronger anticorrelation at $q <$ 30 MeV/c) 
indicates that, on average, protons are emitted later than 
$\alpha$-particles.
Inspection of the  $C_p/C_\alpha$ ratios (Fig.\ \ref{ledpa}, lower panels) 
reveals that the deduced $p\alpha$ emission chronology is qualitatively 
similar for the two beam energies.

\begin{figure}
\vspace{0.4cm}
\centerline{\psfig{file=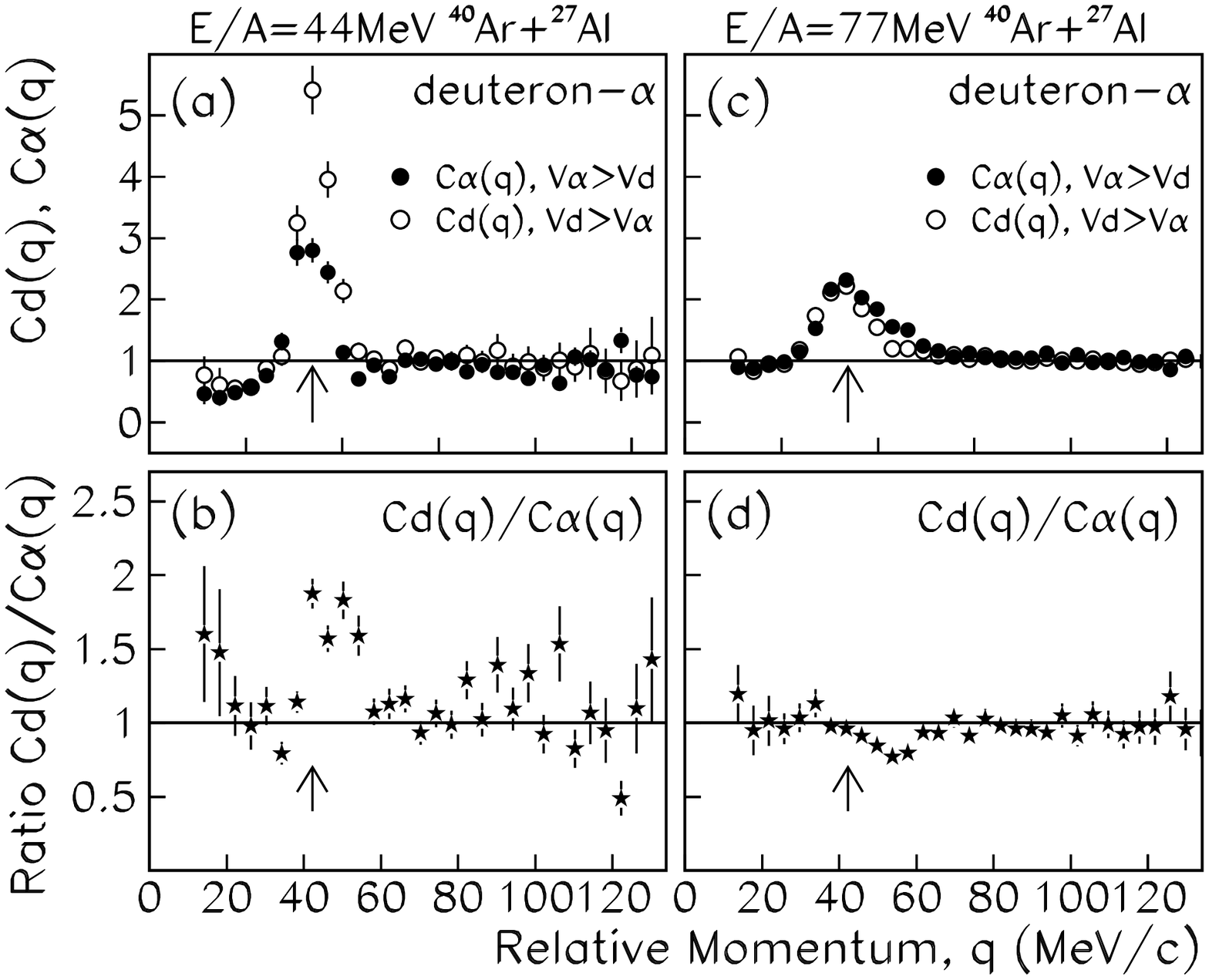,height=12.5cm,angle=0}}
\caption{
From $E/A$ = 44 MeV (left column) and 77 MeV (right column) 
$^{40}$Ar + $^{27}$Al collisions, 
upper panels: particle-velocity-gated 
(filled and open circles) $d\alpha$ correlation functions; 
lower panels: the ratio of the particle-velocity-gated 
correlation functions. 
}
\label{ledda}
\vspace{0.4cm}
\end{figure}

\subsection{The particle-velocity-gated $d\alpha$ correlation function}
The particle-velocity-gated $d\alpha$ 
correlation function is shown in Fig.\ \ref{ledda}. 
For $d\alpha$ pairs, the behavior of the  $C_d/C_\alpha$ ratio 
(Fig.\ \ref{ledda} lower panels) is of delicate interpretation, 
being dominated by the two-body resonance decays of $^6$Li. 
For $d\alpha$ pairs originating from the two-body decay of $^6$Li 
excited states, the inequality $v_d > v_\alpha$ should  
hold in the $^6$Li rest system (due to momentum conservation), 
leading to an enhancement of the $C_d$ correlation 
function in the PLF system at $q \approx$ 42.2 MeV/c 
(as discussed above).
 
The experimental data display such an enhancement of the $C_d$ 
correlation function in the $E/A$ = 44 MeV measurement [Fig.\ \ref{ledda}(a)], 
therefore it is difficult to deduce the emission chronology in this case. 
To this end, one has to rely on the behavior of the 
particle-velocity-gated correlation function 
at $q <$ 30 MeV/c, where there are no resonant states and where the $d\alpha$ 
correlation function displays an anticorrelation \cite{Boal2}. 
However, the statistics is very low in this region.
Nevertheless, the fact that pairs with $v_\alpha > v_d$ interact more strongly 
[the ratio $C_d/C_\alpha$ is larger than unity in this anticorrelation region, 
Fig.\ \ref{ledda}(b)] gives an indication that deuterons are, on average, emitted 
earlier than $\alpha$-particles. 

The kinematical signature of two-body resonance decay is much 
weaker in the $E/A$ = 77 MeV  $d\alpha$ data. 
The enhancement of $C_d$ over $C_\alpha$ is not observed [Fig.\ \ref{ledda}(c)], 
and the ratio $C_d/C_\alpha$ is close to unity at $q \approx$ 42.2 MeV/c 
[Fig.\ \ref{ledda}(d)], while it exhibits a dip in the 
region $q \approx$ 40--70 MeV/c, where the $d\alpha$ correlation function 
displays a correlation \cite{Boal2}. 
Such a dip cannot originate from the decay of $^6$Li 
(for momentum conservation reasons, see previous discussion). 
This indicates that the yield of $d\alpha$ pairs measured at 
$E/A$ = 77 MeV is largely contributed by pairs 
that are not emitted by the decay of $^6$Li excited states, 
but that are emitted independently from PLF or secondary fragment decay. 

The fact that $C_\alpha$ shows a stronger correlation than $C_d$,  
reflects that the average effective distance between the 
two particles is smaller for $C_\alpha$ than for $C_d$. 
Since for $C_\alpha$, $v_\alpha > v_d$, this is only possible 
if (the slower) deuterons are one average emitted before (the faster) 
$\alpha$-particles. Thus, the result of the emission time chronology 
for these pairs, as deduced from the dip in the $C_d/C_\alpha$ ratio 
at $q \approx$ 40--70 MeV/c, is that deuterons are, on average, 
emitted earlier than $\alpha$-particles.

\begin{figure}
\centerline{\psfig{file=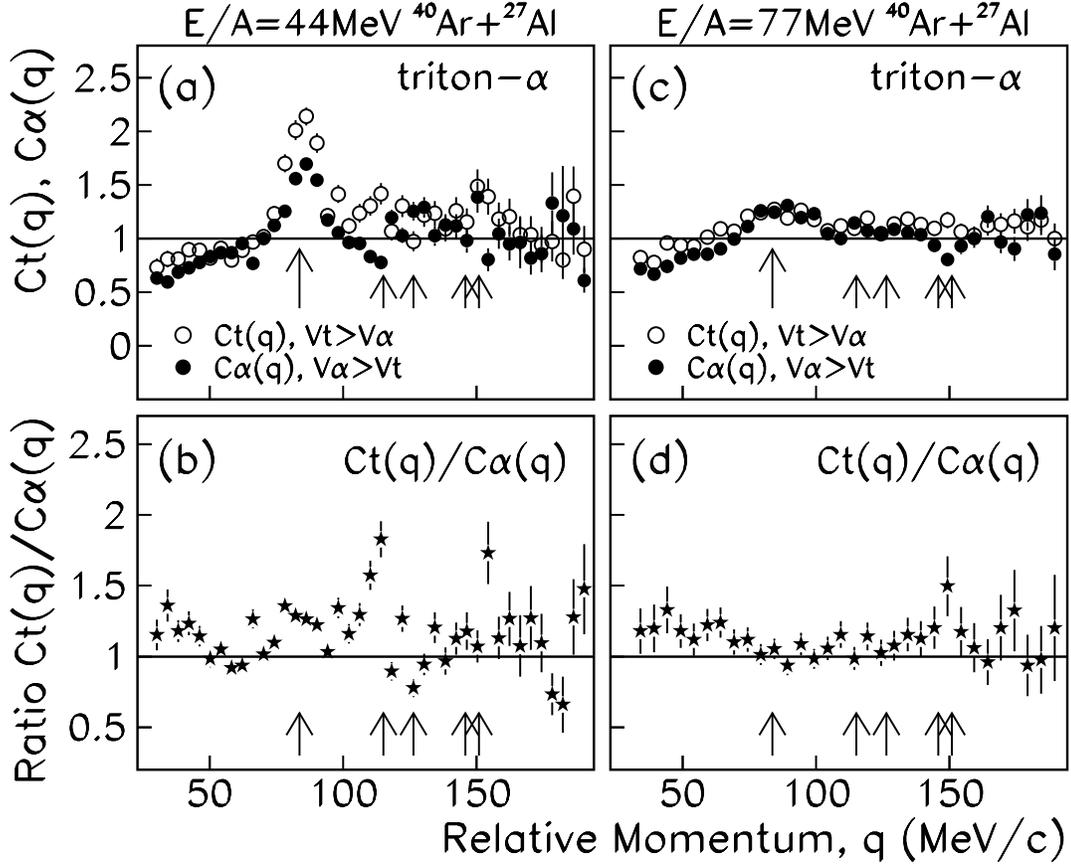,height=12.5cm,angle=0}}
\caption{
From $E/A$ = 44 MeV (left column) and 77 MeV (right column) 
$^{40}$Ar + $^{27}$Al collisions, 
upper panels: particle-velocity-gated (
filled and open circles) $t\alpha$ correlation functions; 
lower panels: the ratio of the particle-velocity-gated 
correlation functions. 
}
\label{ledta}
\vspace{0.4cm}
\end{figure}
\subsection{The particle-velocity-gated $t\alpha$ correlation function}
The particle-velocity-gated 
$t\alpha$ correlation function is shown in Fig.\ \ref{ledta}. 
Also in this case, the resonance peaks are more pronounced 
at $E/A$ = 44 MeV than at 77 MeV. For the $E/A$ = 44 MeV data, 
the behavior of the $C_t/C_\alpha$ ratio at $q >$ 70 MeV/c is dominated 
by the kinematical signature of these two-body decays.
Due to momentum conservation, the inequality $v_t > v_\alpha$ dominates 
in the PLF frame for $t\alpha$ pairs originating from two-body decays, 
leading to an enhancement of the particle-velocity-gated correlation 
function $C_t$. 

The emission chronology is deduced from the behavior of the 
particle-velocity-gated 
correlation function at $q <$ 70 MeV/c, where there are no resonant states 
and the $t\alpha$ correlation function displays an anticorrelation 
(\cite{Boal2} and Fig.\ 2). 
For both $E/A$ = 44 and 77 MeV data, 
this anticorrelation is stronger for $C_\alpha$ 
(i.e.\ the ratio $C_t/C_\alpha$ is larger than unity in this 
anticorrelation region, see Fig.\ \ref{ledta}, lower panels). 
Following the same kind of reasoning discussed above, 
this indicates that tritons are, on average, emitted 
earlier than $\alpha$-particles. 

In summary, the results from the analysis of particle-velocity-gated 
correlation functions indicate that, at both bombarding energies, 
$\alpha$-particles have an average emission time shorter than protons, 
but longer than deuterons and tritons. 
As we have pointed out, because of the dominating behavior of the 
resonant states at $E/A$ = 44 MeV (discussed in Sec.\ \ref{sec:cf}), 
the study of the particle emission chronology from the analysis of 
particle-velocity-gated $d\alpha$ and $t\alpha$ 
correlation functions is rather delicate at $E/A$ = 44 MeV. 
Nevertheless, the indications from the resonance-free, low-$q$ regions, 
support a particle emission time sequence in agreement with that 
deduced at the higher bombarding energy. 

%%%%%%%%%%%%%%%%%%%%%%%%%%%%%%%%%%%%%%%%%%%%%%%%%%%%%%%%%%%%%%%%%%%%%%%%%%%%%%%%%%%%
\section {Summary and conclusions}
\label{sec:summary}

Single particle energy spectra and two-particle correlation 
functions have been measured simultaneously and in coincidence with 
forward emitted fragments from $E/A$ = 44 and 77 MeV $^{40}$Ar + $^{27}$Al 
reactions. The measurements have been performed at 
very forward angles, 0.7$^{\rm o} \le \theta_{lab} \le 7^{\rm o}$, 
with the aim to study the characteristics of 
de-excitation of the PLF source. 

A signature of PLF decay is found in the energy spectra 
of protons, deuterons, tritons and $\alpha$-particles. 
These are characterized by a dumb-bell peak structure, 
originating from those particles, from the isotropically 
emitting PLF, that have a velocity component either parallel 
to PLF movement (higher energy peak) or antiparallel (lower energy peak). 
In order to exclude correlations other than final state interactions 
and quantum symmetrization effects, 
only particles from the higher energy peak have been considered, 
thus enhancing PLF emission \cite{Brief04}. 

The $pp$ correlation function
indicates a lack of final state interaction, 
typical of particles emitted by a long-lived source. 
This feature is found in both $E/A$ = 44 and 77 MeV data, 
pointing to characteristics of the PLF emission source 
that are rather independent of the beam energy, 
for what concerns the long-time tail of the PLF de-excitation chain, 
probed by protons. 

The correlation functions dominated by resonances 
($d\alpha$, $t\alpha$, and $\alpha\alpha$)
consistently display higher resonance peaks 
at $E/A$ = 44 MeV than at 77 MeV. 
This indicates that 
the relative abundance of light particles 
(in particular $d$ and $t$) 
as compared to heavy fragments 
(in particular $^8$Be, $^7$Li, and $^6$Li) 
is larger at $E/A$ = 77 MeV than at $E/A$ = 44 MeV. 

The correlation functions of non-identical particle pairs 
have been utilized to deduce the particle emission 
time sequence in a model independent way, 
based on particle-velocity-gated correlation functions. 

For correlation functions characterized by final state 
interaction leading to resonances, care has to be taken to deduce 
information on the emission time sequence from independently 
emitted particles (discarding particles coming from resonance decays). 
This can be achieved in two ways: \\
i) With particle-velocity gates in regions of relative 
momentum where there are no resonant states that directly 
lead to the formation of the particle pair. \\
ii) With particle-velocity gates in regions of the relative momentum 
where there is a resonance peak, but the peak is not dominated by 
particles coming from resonance decay. When a resonance peak is 
dominated by particles coming from resonance decay, momentum conservation 
implies that the correlation is stronger for the particle-velocity-gated 
correlation function with the velocity of the lighter particle larger 
than the velocity of the heavier particle. Violation of this condition 
is a signature that the peak is dominated by independently emitted 
particles, and can thus be utilized to deduce the particle 
emission time sequence. 

The results from the particle-velocity-gated correlation function 
analysis indicate that, at both bombarding energies, 
$\alpha$-particles have an average emission time shorter than protons but 
longer than deuterons and tritons. 
This finding is consistent with the emerging picture of a PLF that is de-excited 
by early emission of tritons and deuterons 
followed by $\alpha$-particles and protons. 
Indeed, because of the Coulomb barrier and of the binding energy 
relative to the available excitation of the PLF, 
the emission time distribution of protons and  $\alpha$-particles is 
expected to be less steep than that of $d$ and $t$ \cite{Hudan}. 

The deduced LCP emission time sequence appears to be in qualitative 
agreement with previous experimental findings from three-moving 
source fits to single particle energy spectra in $E/A$ = 44 and 60 MeV 
$^{40}$Ar + $^{27}$Al collisions  \cite{Lanza98,Lanza01}. 
The PLF source temperatures deduced from these fits were found 
to be lowest for protons (3.6 MeV at 44 A MeV, 4.9 MeV at 60 A MeV), 
followed by $\alpha$-particles (4.7 MeV at 44 A MeV, 6.9 MeV at 60 A MeV), 
then deuterons (5.0 MeV at 44 A MeV, 8.5 MeV at 60 A MeV) 
and tritons (5.9 MeV at 44 A MeV, 8.5 MeV at 60 A MeV). 
These findings were interpreted as a suggestion that an important 
fraction of $\alpha$-particles is emitted in the final stage of the collision. 
Different hypotheses have been advanced to explain this late emission of 
$\alpha$-particles, including the concept of $\alpha$-clusterization of 
nuclei \cite{Hodgson} and the production of $\alpha$-particles by secondary 
decay of light fragments produced in the collisions \cite{LanzaGanil,LanzaBormio}. 

The qualitative interpretations of the data analysis given in the 
present paper should be supported by comparisons to statistical 
model calculations. Such model comparisons will be presented 
in a forthcoming paper \cite{Helgesson}.
\\

ACKNOWLEDGEMENTS

Financial support from the Swedish Research Council 
(Contracts No.\ F 620-149-2001 and 621-2002-4609), and from 
``Istituto Nazionale di Fisica Nucleare, Sezione di Catania'',
is acknowledged. 

%%%%%%%%%%%%%%%%%%%%%%%%%%%%%%%%%%%%%%%%%%%%%%%%%%%%%%%%%%%%%%%%%%%%%%%%%%%%%%%%%%
%\newpage

\end{document}